\documentclass[fleqn,twoside]{article}
\usepackage{espcrc2}
\usepackage{graphicx}
\usepackage[figuresright]{rotating}

\newcommand{\AmS}{{\protect\the\textfont2
  A\kern-.1667em\lower.5ex\hbox{M}\kern-.125emS}}
\title{Splitting Neutrino masses and Showering into Sky}

\author{D.Fargion\address[roma1]{Physics Department, Rome University La Sapienza,
00185, Ple.A.Moro 2, Rome, Italy.\\ $^{\dag}$ INFN, Istituto
Nazionale di Fisica Nucleare.}$^{\dag}$\thanks{The author thanks
Drs. A. Colaiuda.},
      D. D'Armiento\addressmark[roma1],
      O. Lanciano\addressmark[roma1],
      P. Oliva\addressmark[roma1],
      M. Iacobelli\addressmark[roma1],
      P.G. De Sanctis Lucentini\addressmark[roma1],
      M. Grossi\addressmark[roma1],
      M. De Santis\addressmark[roma1].}

\begin{document}

\begin{abstract}
Neutrino masses might be as light as a few time the atmospheric
neutrino mass splitting. The relic cosmic neutrinos may cluster in
wide Dark Hot Local Group Halo. High Energy ZeV cosmic neutrinos (in
Z-Showering model) might hit relic  ones  at each mass in different
resonance energies in our nearby Universe. This non-degenerated
density and energy must split UHE Z-boson secondaries (in Z-Burst
model) leading to multi injection of UHECR nucleons within future
extreme AUGER energy. Secondaries of Z-Burst  as neutral gamma,
below a few tens EeV are better surviving local GZK cut-off and they
might explain recent Hires BL-Lac UHECR correlations at small
angles. A different high energy resonance must lead to Glashow's
anti-neutrino showers while hitting electrons in matter. In water
and ice it leads to isotropic light explosions. In air, Glashow's
anti-neutrino showers lead to collimated and directional air-showers
offering a new Neutrino Astronomy. Because of neutrino flavor
mixing, astrophysical energetic tau neutrino above tens GeV must
arise over atmospheric background. At TeV range is difficult to
disentangle tau neutrinos from other atmospheric flavors. At greater
energy around PeV, Tau escaping mountains and Earth and decaying in
flight are effectively showering in air sky. These Horizontal
showering is splitting by geomagnetic field in forked shapes. Such
air-showers secondaries release amplified and beamed gamma bursts
(like observed TGF), made also by muon and electron pair bundles,
with their accompanying rich Cherenkov flashes. Also planet's
largest (Saturn, Jupiter) atmosphere limbs offer an ideal screen for
UHE GZK and Z-burst tau neutrino, because their largest sizes. Titan
thick atmosphere and small radius are optimal for discovering
up-going resonant Glashow resonant anti-neutrino electron showers.
Detection from Earth of Tau, anti-Tau, anti-electron neutrino
induced Air-showers by twin Magic Telescopes on top mountains, or
space based detection on balloons and satellites arrays facing the
atmosphere's limbs are the simplest and cheapest way toward UHE
Neutrino Astronomy Horizons.
\end{abstract}

\maketitle

\section{Introduction: $\nu_{\tau}$, $\overline{\nu}_{\tau}$, ${\overline{\nu}_e}$  Astronomy}
Neutrino astronomy by its $\tau$ flavor has been noted only recently
\cite{Fargion1999,Fargion2002a,Athar,Feng2002}, so called HorTaus or
Earth skimming neutrinos.
\begin{figure}[!t]
\includegraphics*[width=70mm,height=60mm,angle=0]{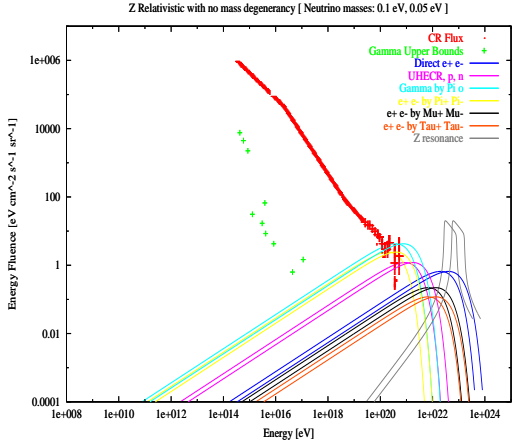}
\vskip-28pt
 \caption{\footnotesize{Twin Z-Burst Showering \cite{Fargion-Mele-Salis99}
 occurs at different peak energy (resonance energies $E_{\nu_i}=M_Z^2/2\cdot
m_{\nu_i}\simeq4\cdot10^{22}(\frac{m_{\nu_i}}{0.1 eV})^{-1}$ eV,)
for non degenerated neutrino masses, as $0.1$ and $0.05$ eV; its
consequent secondary spectra ($\gamma$, nucleons, electron pairs)
inject ZeV UHECR \textit{(p, n)}$
E_p\simeq8\cdot10^{20}\cdot(m_{\nu_i}/0.1 eV)^{-1}$ eV, and
$\gamma$, $E_{\gamma} \simeq$ a few
$10^{19}\cdot(\frac{m_{\nu_i}}{0.1 eV})^{-1}$ eV (by $\pi^0$).
\cite{Fargion2005}}}\label{fig1}
 \end{figure}
 \begin{figure}[!ht]
\includegraphics*[width=60mm,height=80mm, angle=270]{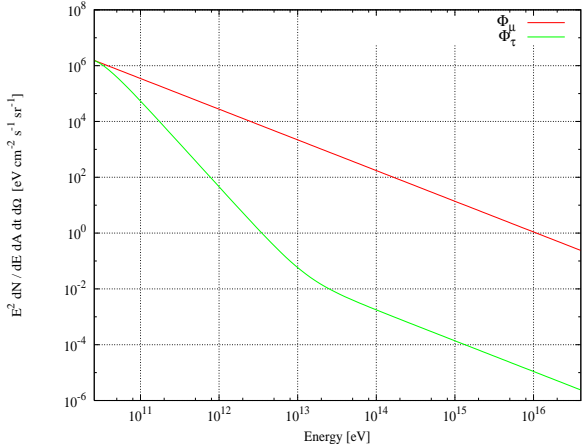}
\vskip-25pt
 \caption{\footnotesize{The  $\nu_{\mu}$, $\overline{\nu}_{\mu}$ is mixed and converted into
  $\nu_{\tau}$, $\overline{\nu}_{\tau}$ with suppressions at high
  energy, making tau neutrino astronomy noise-free over atmospheric
  signals. Here an approximate ${\phi}_{\nu_{\mu}}\simeq{E_{\nu_{\mu}}}^{-3.2}$
  power law has been assumed for atmospheric and prompt neutrinos.
  An astrophysical (Waxman-Bachall\cite{WB97}) UHE fluence  ${\phi}_{\nu}\cdot{E_{\nu}}^{2}\simeq30$ eV $cm^{-1}\,sr^{-1}\,s^{-1}$
  might rule above atmospheric noise already at TeVs energies for $\nu_{\tau}$,
 $\overline{\nu}_{\tau}$, while only at PeVs ones for $\nu_{\mu}$, $\overline{\nu}_{\mu}$ polluted  signals. \cite{Athar}}}\label{fig2}
 \end{figure}
 \begin{figure}[!ht]
 \includegraphics*[width=75mm,height=50mm,angle=0]{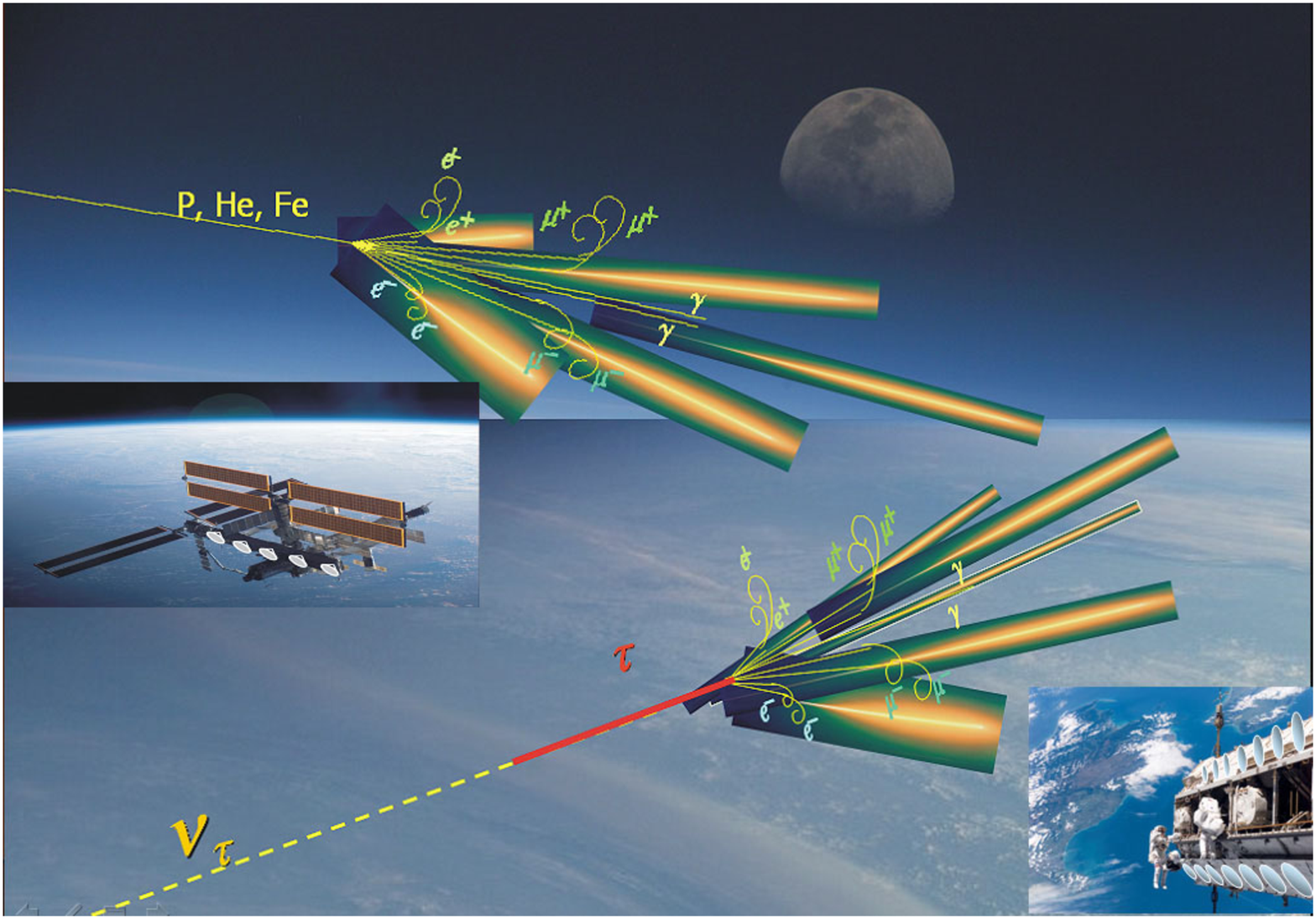}
\vskip-20pt
 \caption{\footnotesize{The forked showering at
 horizons by \textit{UHECR} or $\tau$: their five channels (mostly $\gamma$, electron and $\mu$ pairs) are typical of the hadronic
 event. The fan-shaped jets arise because of the geo-magnetic bending of charged particles at high quota ($\sim23-40$ km).
 These Showers are hundreds km long whose signature would be discovered by Space Station Arrays (Optical, $\gamma$, $e^\pm$, $\mu^\pm$)
 facing the Earth Limb or EUSO looking downward. UHECR secondaries in horizontal showers below the horizons are mostly made by $\mu^\pm$ pairs.}}\label{fig3}
 \includegraphics[width=75mm,height=45mm,angle=0]{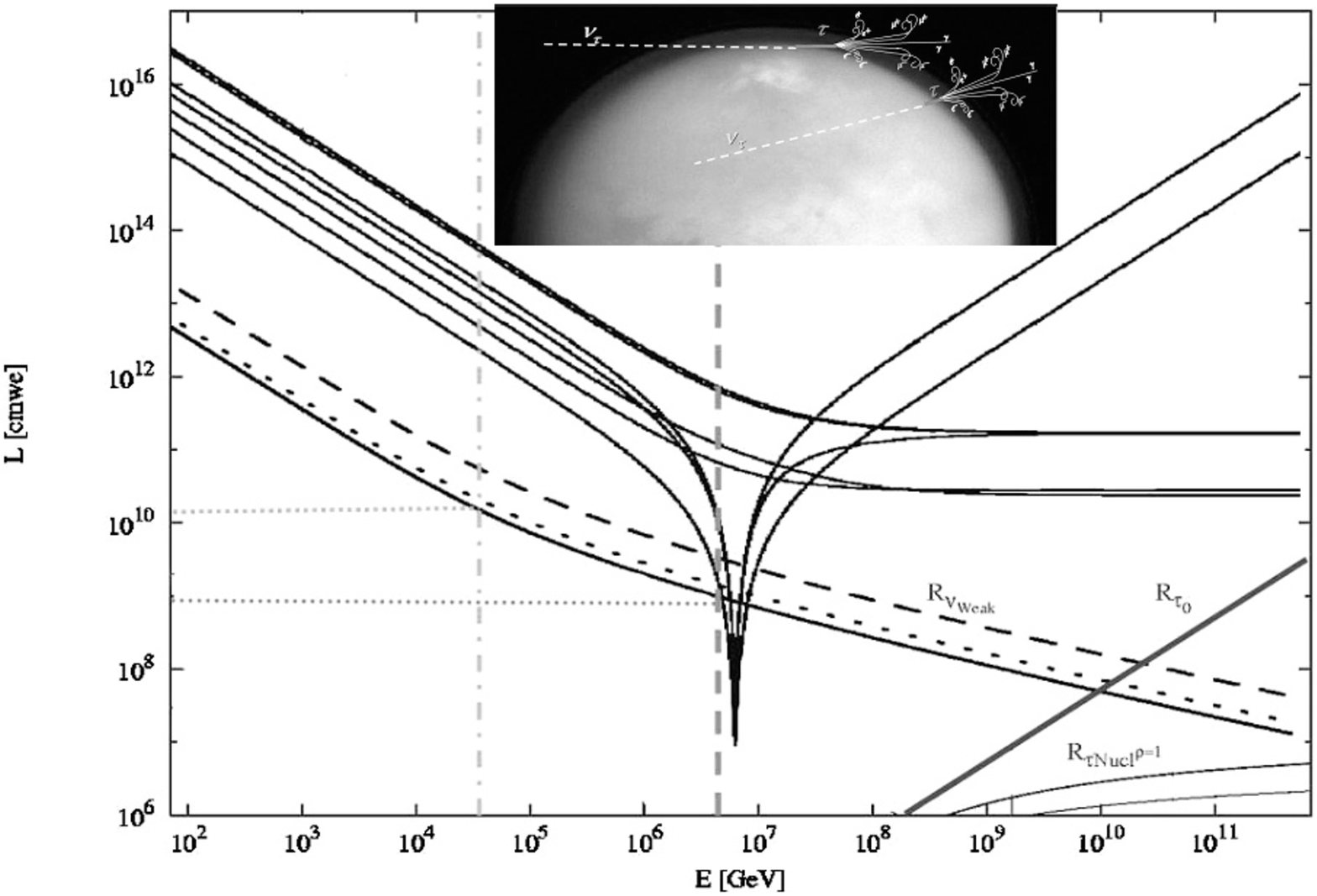}
\vskip-29pt
 \caption{\footnotesize{The interaction length for UHE neutrinos with nuclei and $\overline{\nu}_{e}$ with electron as a
 function of their energy, crossing the Earth (higher dashed line), at $10^{10}$ cmwe, and Titan smaller size
 satellite \cite{Gandhi98,Fargion2005}. Because of it up-going Glashow resonant $\overline{\nu}_{e}$
  cannot cross vertically and air-shower from the Earth, but they may from Titan atmosphere.\cite{Fargion2005} }}\label{fig4}
 \end{figure}
 \begin{figure}[h]
 \vskip-10pt
\includegraphics[width=80mm,height=42mm,angle=0]{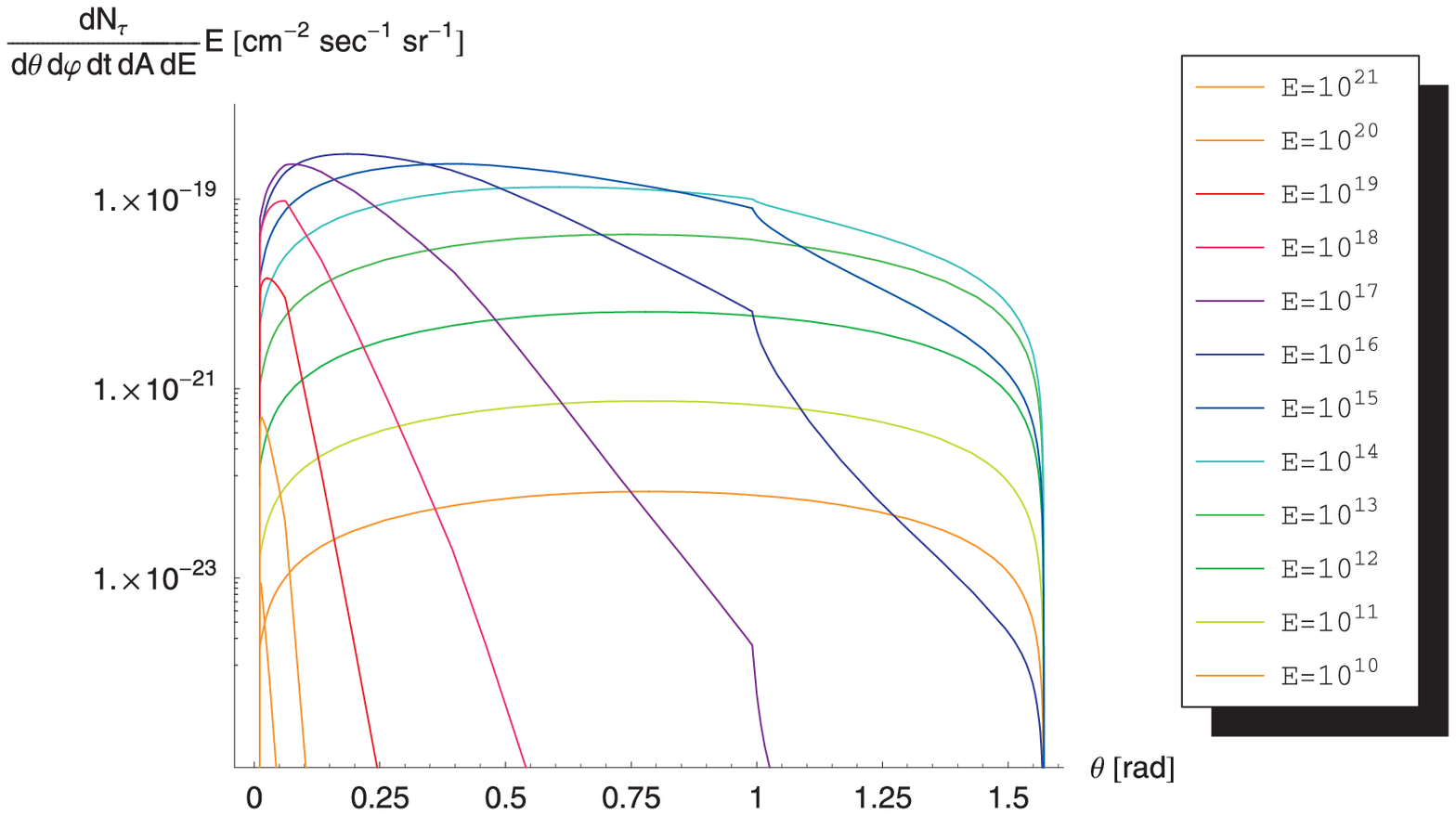}
\includegraphics[width=80mm,height=50mm,angle=0]{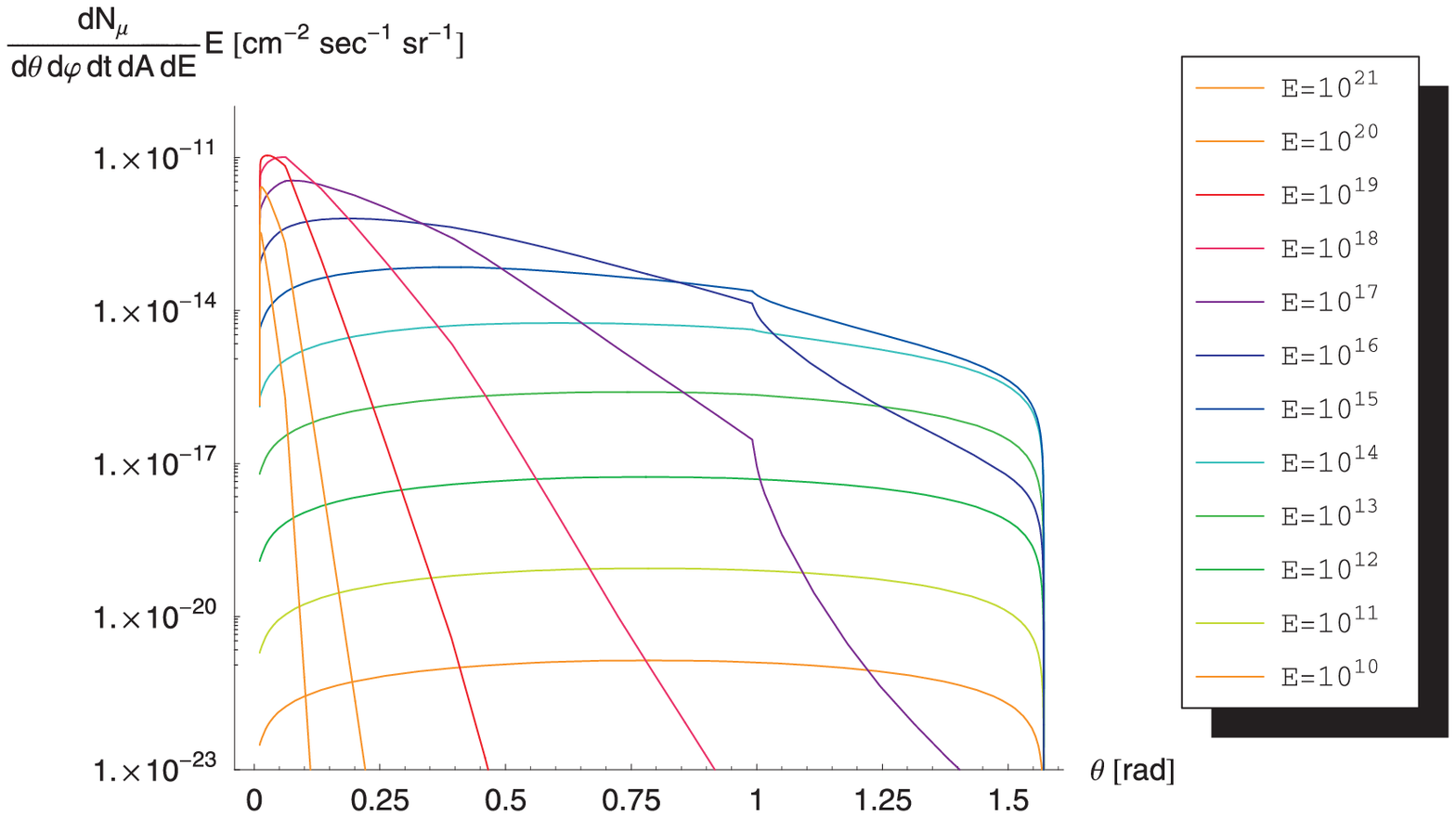}
\includegraphics[width=80mm,height=58mm,angle=0]{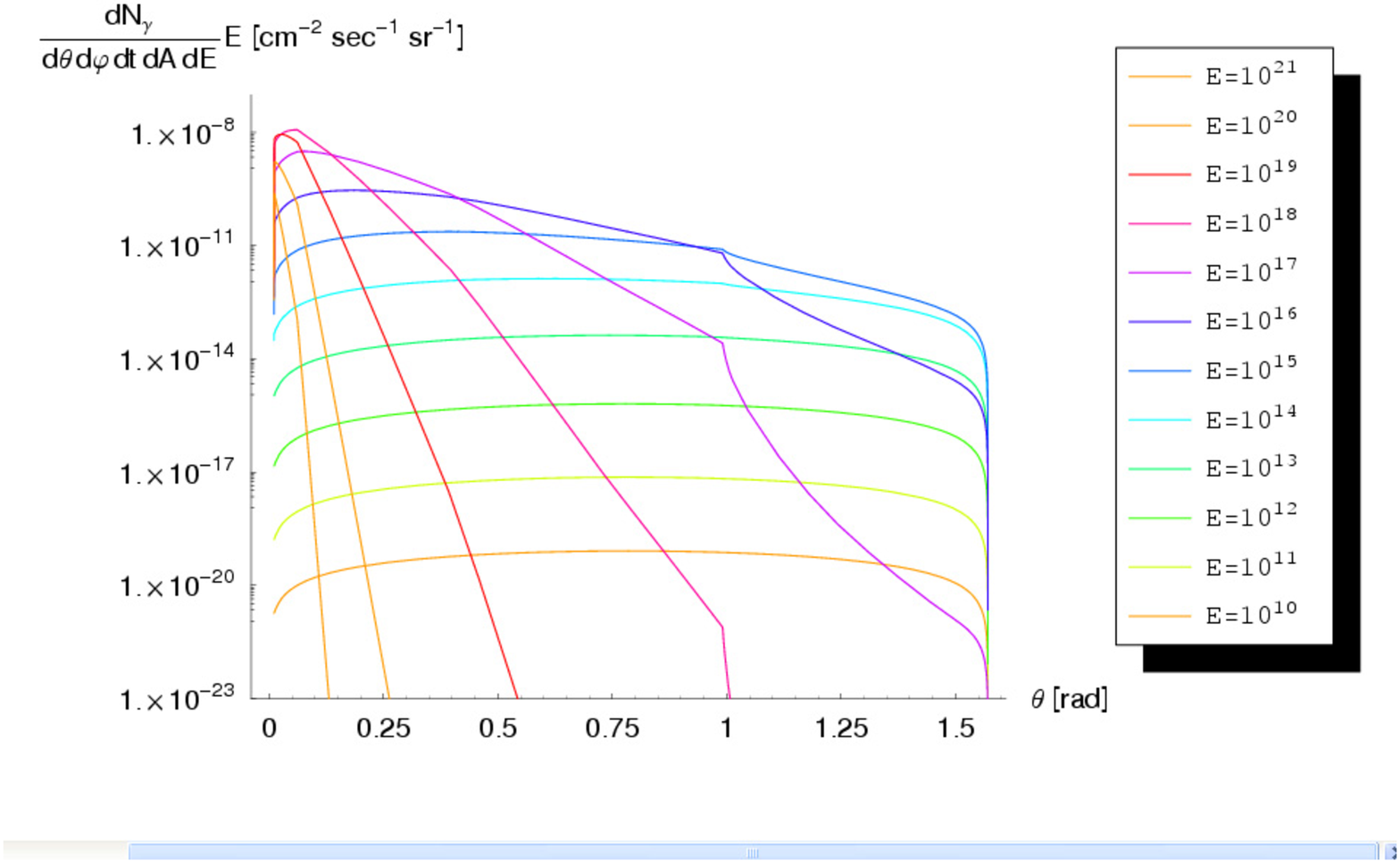}
\vskip-10pt
 \caption{\footnotesize{Up: The average differential
number of event rate of $\tau$ leptons (HorTaus) for an input GZK
constant neutrino energy fluency
${\phi}_{\nu_{\mu}}\cdot{E_{\nu_{\mu}}}^{2}\simeq30$ eV
$cm^{-1}\,sr^{-1}\,s^{-1}$ at Waxman-Bachall \cite{WB97} level. We
are assuming that $\tau$'s are escaping from an Earth outer layer
made of rock \cite{Fargion2004,Fargion2004b}. Note the
discontinuity at $\theta\simeq1$ rad, due to terrestrial higher
density core profile. The smallest homogeneous curve correspond to
lowest UHE neutrino  from $10^{10}$ to $10^{21}$ eV energy.
Middle: The consequent differential number of event rate of the
secondary muons (exceeding at horizons the upward atmospheric
induced ones), as before figure. Last: $\gamma$ flux produced by
the decay in flight of $\tau$ leptons in the Earth's atmosphere. A
proportional parasite Cherenkov photons ($\simeq10^4$ more
abundant than $\gamma$ ones) from these air-showers behave with a
comparable angular profile
\cite{Fargion2004,Fargion2004b,Fargion2005}.}}\label{fig5}
\end{figure}

Neutrino masses, their splitting and mixing, see Fig.\ref{fig1},
lead to a $\nu_{\tau}$ rise. Indeed $\nu$ mixing guarantees a
complete oscillation even at high energy, because of modest (in
cosmic unity) neutrino oscillation distance:
$$
L_{\nu_{\mu}\rightarrow\nu_{\tau}}=32\,pc\biggl(\frac{E_{\nu}}{10^{21}\,eV}\biggr)\biggl({\frac{\Delta
m_{ij}^2}{2.5 \cdot 10^{-3}\,eV ^{2}}}\biggr)^{-1}
$$
enriching rare $\nu_{\tau}$ lepton components. For an opposite
argument \cite{Athar} at energies above $12.9$ GeV up-going
atmospheric $\mu$ neutrino are poorly converted into $\tau$ ones,
see Fig.\ref{fig2}. Therefore already at TeVs energies  and above,
$\nu_{\tau}$ Astronomy is noise free. But only at higher energies,
$E_{\nu_{\tau}}\geq$~PeV the $\nu_{\tau}$ air showering is well
detectable \cite{Fargion2002a}. The problem of $\tau$ neutrinos
crossing, interacting external crust and escaping, as UHE $\tau$,
the Earth \cite{Fargion2004} is  complicated:  the terrestrial
neutrino opacity at different energies and angles of arrival, the
$\tau$ energy losses and its interaction lengths at different
energies and materials, makes difficult to estimate the expected
rate. Such a prediction is further complicated by the existence of
a list of theoretical models for the assumed incoming neutrino
fluxes (GZK neutrinos, Z-burst model flux (See Fig.\ref{fig1} and
\cite{Fargion2004}), $E^{-2}$ flat spectra, AGN neutrinos,
topological defects). Even extreme light and non-degenerated
neutrino masses may play a role in UHECR spectra split in multi
bump and ZeV energy modulations; see Fig \ref{fig1}. Most of the
authors focus on the UpTaus tracks in underground $km^3$
detectors.  We showed that our results give an estimate of the
$\tau$ air-shower event rates that exceeds earliest studies
\cite{Bottai2002} but they are comparable or even below more
recent predictions \cite{Yoshida2004}. Different rate estimations
are leading to a vanishing number of events a year
\cite{Bottai2002} in EUSO,  a fraction of unity of them
\cite{Bertou2002} or a few of events a year \cite{Fargion2004} in
AUGER array observable from the Ande shadows (on West side) as
well as  a dozen of events for EUSO in three years
\cite{Fargion2004}. We pointed out \cite{Fargion2004,Fargion2004b}
that the consequent $\mu^{\pm}$, $e^{\pm}$, $\gamma$  bundle
fluence and signature of HorTaus largely differs (by its
electromagnetic richness) from that of horizontal filtered UHECR
backgrounds at far horizons. Also new net charge single bundle (by
HorTaus) are possible at high altitude splitting air-showers (see
Fig.\ref{fig3})\cite{Fargion2004,Fargion2004b}. We apply all our
previous results to the calculation of the averaged expected
number of $\mu^{\pm}$, $e^{\pm}$ and $\gamma$ in the up atmosphere
see Fig.\ref{fig5}. These up-going muon peak flux
$\simeq10^{-11}\,cm^{-2}s^{-1}sr^{-1}$ ($\simeq 6^o$ below
horizons), originated by \textit{guaranteed} GZK $\tau$
secondaries, exceed the observed up-going muon atmospheric one
$\simeq10^{-13}\,cm^{-2}s^{-1}sr^{-1}$ (induced by atmospheric
neutrinos), making the search of rare bundles a realistic Neutrino
telescope above the noise. Crown Array by large areas surfaces
($10^2-10^3$ $m^2$) on top mountains are needed
\cite{Fargion2004b}. This search at horizons atmosphere regards
also Glashow's resonance $\overline{\nu}_e+e\rightarrow
W^-\rightarrow X$, $E_{\overline{\nu}_e}= \frac{{M_W}^2}{2\cdot
m_{e}}\simeq6.4\cdot 10^{15}$ eV. Also $\tau$-channel may lead via
$\overline{\nu}_e+e\rightarrow
W^-\rightarrow\tau+\overline{\nu}_{\tau}$ to Tau air-showers.
Similar to $\overline{\nu}_e+e\rightarrow W^-\rightarrow X$
neutrino astronomy might arise in SUSY models by UHE $\chi_0$ via
s-electrons resonance
$\chi_0+e\rightarrow\tilde{e}\rightarrow\chi_0+e$
\cite{Datta2004}. These showers will split  by geomagnetic fields,
into  twin fingers defined by their charges (muon and electron
pair arcs and rectilinear $\gamma$ paths for tens and hundreds of
km distances, see (Fig.\ref{fig3}). Their imprint may be soon
revealed in on going Magic Telescope test at horizons by the
discover of twin spots lights a few degree one from the
other,\cite{Fargion2005}. Also up-ward horizontal muon bundles at
top mountain in present (USA) and future Milagro detectors
(Bolivia) are a very promising road to  astrophysical
${\nu}_{\tau}$,${\overline{\nu}_e}$ neutrino air-showers. In
conclusion, neutrino astronomy may soon shine at horizons of Earth
by air-showers seen in space (as from satellites and balloons
arrays) or from mountains: their signatures maybe soon found in
\textit{Magic} Cherenkov split lights detections by the canaries
twin telescope facing the far horizons, within muon and gamma
bundle Cherenkov rings, in time and direction coincidence with
BL-Lac, Mrk, GRB rising sources
\cite{Fargion2004,Fargion2001a,Fargion2005}.


\begin{thebibliography}{100}
\bibitem{Fargion1999}{\normalsize D.Fargion, A.Aiello et al. 26th ICRC, He 6.1.10 (1999)
396-398}.
\bibitem{Fargion2002a}{\normalsize D.Fargion, ApJ  570 (2002) 909-925; astro-ph/0002453;
astro-ph/9704205}.
\bibitem{Athar}{\normalsize  H.Athar,  Chin.J.Phys. 42 (2004) 1-20 \textit{and} D.Fargion, Phys.Scripta T127 (2006)
22-24}.
\bibitem{Feng2002}{\normalsize J.L.Feng, et. al., Phys.Rev.Lett. 88 (2002) 161102;
hep-ph/0105067}.
\bibitem{Fargion2004}{\normalsize D.Fargion et al. ApJ 613 (2004) 1285-1301; hep-ph/0305128; astro-ph/0607526 \textit{and} astro-ph/0501033; \textit{and} astro-ph/0501079 in Adv.Space Res. 37 (2006)
2132-2138}.
\bibitem{Bottai2002}{\normalsize S.Bottai, S.Giurgola, Astrop.Phys. BG03 18 (2003)
539}.
\bibitem{WB97}{\normalsize E.Waxman,  J.Bachall, Phys.Rev.Lett. 78 (1997)
2292}.
\bibitem{Yoshida2004}{\normalsize  S.Yoshida et al., Phys.Rev. D69 (2004) 103004;
astro-ph/0312078}.
\bibitem{Bertou2002}{\normalsize  X.Bertou et al., Astropart.Phys.,  17 (2002)
183}.
\bibitem{Fargion2004b}{\normalsize D.Fargion et al., Nucl.Phys.B (Proc.Suppl.) 136 (2004) 119;
astro-ph/0409460}.
\bibitem{Gandhi98}{\normalsize R.Gandhi et al., Phys.Rev.D., 58 (1998)
093009}.
\bibitem{Fargion2005}{\normalsize D.Fargion, Prog.Part.Nucl.Phys. 57 (2006) 384-393; \textit{and} D.Fargion et al. Nucl.Phys.B (in press 2007); astro-ph/0610725,
astro-ph/0610954}.
\bibitem{Datta2004}{A.Datta, D.Fargion, B.Mele,  JHEP 0509 (2005)
007}.
\bibitem{Fargion2001a}{\normalsize D.Fargion, Proc. 27th ICRC (Hambourg) HE2.5 (2001) 1297-1300;
astro-ph/0106239}.

\bibitem{Fargion-Mele-Salis99}{\normalsize D.Fargion, B.Mele, A.Salis,  ApJ 517 (1999) 725;
astro-ph/9710029}.


\end{thebibliography}
\end{document}